\documentclass[12pt,preprint]{aastex}
\usepackage{epsfig}
\usepackage{graphicx}
\usepackage{slashbox}
\usepackage{multirow}
\usepackage{lscape}
\usepackage{mathrsfs,amssymb}
\usepackage{amsmath}

\newcommand       \mum        {\,{\rm \mu m}}

\newcommand       \simali       {{\sim}\,}
\newcommand       \magni        {\,{\rm mag}}
\newcommand       \simlt        {\lesssim}
\newcommand       \simgt        {\gtrsim}
\newcommand       \gtsim        {\gtrsim}

\newcommand       \Angstrom     {\,{\rm \AA}}
\newcommand       \g            {\,{\rm g}}
\newcommand       \cm           {\,{\rm cm}}

\newcommand       \s            {\,{\rm s}}
\newcommand       \erg          {\,{\rm erg}}
\newcommand       \nH           {n_{\rm H}}
\newcommand       \NH           {N_{\rm H}}
\newcommand       \rmH          {\,{\rm H}}

\newcommand       \K            {\,{\rm K}}
\newcommand       \mH         {m_{\rm H}}

\newcommand{\RV}{R_{\rm V}}

\newcommand       \cgas     {\left[{\rm C/H}\right]_{\rm gas}}

\newcommand       \sidust   {\left[{\rm Si/H}\right]_{\rm dust}}
\newcommand       \cdust    {\left[{\rm C/H}\right]_{\rm dust}}

\newcommand \cPAH {\left[{\rm C/H}\right]_{\scriptscriptstyle\rm PAH}}
\newcommand \cVLG {\left[{\rm C/H}\right]_{\scriptscriptstyle\rm VLG}}
\newcommand \siVLG {\left[{\rm Si/H}\right]_{\scriptscriptstyle\rm VLG}}
\newcommand \feVLG {\left[{\rm Fe/H}\right]_{\scriptscriptstyle\rm VLG}}

\newcommand       \Cs      {C_{\rm s}}
\newcommand       \Cg      {C_{\rm g}}
\newcommand       \alphas  {\alpha_{\rm s}}
\newcommand       \alphag  {\alpha_{\rm g}}
\newcommand       \betas   {\beta_{\rm s}}
\newcommand       \betag   {\beta_{\rm g}}
\newcommand       \ats     {a_{\rm t,s}}
\newcommand       \atg     {a_{\rm t,g}}
\newcommand       \acs     {a_{\rm c,s}}
\newcommand       \acg     {a_{\rm c,g}}

\newcommand       \ppm     {\,{\rm ppm}}

\newcommand       \bVLG    {b_{\scriptscriptstyle\rm VLG}}
\newcommand \siism {\left[{\rm Si/H}\right]_{\scriptscriptstyle\rm ISM}}

\newcommand \sisun {\left[{\rm Si/H}\right]_\odot}
\newcommand \csun  {\left[{\rm C/H}\right]_\odot}

\newcommand \sistar {\left[{\rm Si/H}\right]_\star}
\newcommand \cstar  {\left[{\rm C/H}\right]_\star}

\shorttitle{Very Large Interstellar Grains}
\shortauthors{Wang, Li \& Jiang}

\begin{document}

\title{
 \vspace*{-2.0em}
  {\normalsize\rm Accepted for publication in 
                  {\it The Astrophysical Journal}}\\
 \vspace*{1.0em}
Very Large Interstellar Grains as Evidenced
by the Mid-Infrared Extinction\footnote{%
  Dedicated to the late Professor
  J.~Mayo~Greenberg (1922.1.14--2001.11.29)
  of Leiden University
  who first suggested the possible existence
  of very large grains in the interstellar space.
  }
  }
\author{Shu Wang\altaffilmark{1,2},
        Aigen Li\altaffilmark{2}, and
        B.W.~Jiang\altaffilmark{1}}
\altaffiltext{1}{Department of Astronomy,
                 Beijing Normal University,
                 Beijing 100875, China;
                 {\sf shuwang@mail.bnu.edu.cn,
                      bjiang@bnu.edu.cn}
                 }
\altaffiltext{2}{Department of Physics and Astronomy,
                 University of Missouri,
                 Columbia, MO 65211, USA;
                 {\sf wanshu@missouri.edu,
                       lia@missouri.edu}
                 }

\begin{abstract}
The sizes of interstellar grains
are widely distributed,
ranging from a few angstroms to a few micrometers.
The ultraviolet (UV) and optical extinction
constrains the dust in the size range of
a couple hundredth micrometers
to several submicrometers.
The near and mid infrared (IR) emission
constrains the nanometer-sized grains
and angstrom-sized very large molecules.
However, the quantity and size distribution
of micrometer-sized grains remain unknown
as they are gray in the UV/optical extinction
and they are too cold and emit too little
in the IR to be detected by
{\it IRAS}, {\it Spitzer}, or {\it Herschel}.
In this work,
we employ the $\simali$3--8$\mum$ mid-IR extinction
which is flat in both diffuse and dense regions
to constrain the quantity, size, and composition
of the $\mu$m-sized grain component.
We find that, together with nano- and submicron-sized
silicate and graphite (as well as PAHs),
$\mu$m-sized graphite grains with
C/H\,$\approx$\,137$\ppm$ and a mean size
of $\simali$1.2$\mum$ closely fit
the observed interstellar extinction
of the Galactic diffuse interstellar medium
from the far-UV to the mid-IR
as well as the near-IR
to millimeter thermal emission
obtained by {\it COBE}/DIRBE,
{\it COBE}/FIRAS, and {\it Planck}
up to $\lambda\simlt1000\mum$.
The $\mu$m-sized graphite component accounts for
$\simali$14.6\% of the total dust mass
and $\simali$2.5\% of the total IR emission.
\end{abstract}


\section{Introduction}
The absorption and scattering
--- their combination
is called ``extinction'' ---
of starlight are caused by interstellar grains
of all sizes. But grains of sizes comparable to
the wavelength ($\lambda$) of starlight absorb
and scatter photons most effectively
(i.e., $2\pi a/\lambda\sim1$, where $a$
is the spherical radius of the grain; see Li 2009).
Because of this, interstellar grains have
long been known to be ``submicron-sized'',
with a canonical size of $\simali$0.1$\mum$
(i.e., $a$\,$\sim$\,$\lambda_V/2\pi$\,$\simali$0.1$\mum$
for the visual band $\lambda_V=5500\Angstrom$)
since their first detection via extinction
and reddening in the visible (Trumpler 1930).
However, it is now well recognized that
they actually span a range of sizes
from 
subnanometers (i.e., angstroms)
and nanometers to submicrometers and micrometers.

The sub-$\mu$m-sized grain population is well
constrained by the wavelength-dependent extinction
from the near infrared (IR) at $\lambda<3\mum$
to the far ultraviolet (UV) at $\lambda>0.1\mum$.
This population, ranging from
a couple hundredth micrometers
to several submicrometers,
is often referred to as
``sub-$\mu$m-sized'' grains or ``classical'' grains.

The angstrom- and nano-sized grain population cannot be
constrained even by the far-UV extinction.
These grains
(of sizes from $a<50\Angstrom$ down to a few angstroms)
are often referred
to as ``very large molecules'',
``very small grains'' (VSG),
``ultrasmall grains'',
or ``nanoparticles''.
They are in the Rayleigh regime
(i.e., $2\pi a/\lambda\ll1$) in the far-UV.
On a per unit volume ($V$) basis,
their extinction cross section
$C_{\rm ext}(a,\lambda)$ of spherical radii $a$
at wavelength $\lambda$
is independent of grain size, i.e.,
$C_{\rm ext}(a,\lambda)/V = C_{\rm ext}(\lambda)/V$.
The extinction resulting from
these grains is not sensitive to
the size distribution $dn/da$
but their total volume:
\begin{eqnarray}
A_\lambda/\NH & = &
         1.086\,\int da\,\frac{1}{\nH}\frac{dn}{da}\,
         C_{\rm ext}(a,\lambda)
         \nonumber\\
 & &   = 1.086\,\frac{C_{\rm ext}(\lambda)}{V}\,
         \int da\,\frac{1}{\nH}\frac{dn}{da}\,
         \frac{4\pi}{3}a^3
         \nonumber\\
 & &   = 1.086\,\frac{C_{\rm ext}(\lambda)}{V}\,
         \left(\frac{V_{\rm VSG}}{\rm H}\right) ~,
\label{eq:nano}
\end{eqnarray}
where $\nH$ ($\NH$) is 
the hydrogen volume (column) density,
and $V_{\rm VSG}/{\rm H}$ is the total volume
per H nuclei of these grains.
Eq.\,\ref{eq:nano} demonstrates that
the observed far-UV extinction
is not able to constrain
the size distribution of this grain population.
On the other hand,
with a heat content smaller or comparable to
the stellar photons that heat them,
these grains are stochastically heated
by single photons
and do not attain an equilibrium temperature
(Draine \& Li 2001).
Upon absorption of an energetic photon
of energy $h\nu$, a ultrasmall grain will be
heated to a maximum temperature $T_{p}$ determined
by its specific heat $C(T)$ and $h\nu$:
$\int_{0}^{T_{p}} C(T)\,dT = h\nu$,
and then rapidly radiates away most of
the absorbed energy at temperature $T_{p}$.
Therefore, its IR emission contains crucial
information about its size
since $C(T)\propto a^3$ (see Li 2004).
Indeed, as demonstrated by Li \& Draine (2001; LD01)
and Draine \& Li (2007; DL07),
the size distribution of
the ultrasmall grain population
is determined by the near- and mid-IR emission
features at 3.3, 6.2, 7.7, 8.6 and 11.3$\mum$
ascribed to polycyclic aromatic hydrocarbon (PAH),
as well as the near- and mid-IR broadband photometry
of {\it IRAS} and {\it COBE}/DIRBE,
particularly those at $\lambda<25\mum$.

In the UV/optical, micrometer-sized grains
or ``very large grains'' (VLG)
are in the geometrical-optics regime
(i.e., $2\pi a/\lambda\gg 1$)
and their extinction 
is essentially wavelength-independent
or ``gray'' (Li 2009).
Therefore, the UV/optical extinction
is not able to constrain their quantity
or size distribution.
There are several lines of direct evidence
for the presence of very large grains
in the interstellar medium (ISM).
Micrometer-sized interstellar
SiC, graphite, Al$_2$O$_3$,
and Si$_3$N$_4$ grains
have been found
in primitive meteorites.
They were identified as presolar grains
of stellar origin based on
their isotope anomalies (Clayton \& Nittler 2004).
Furthermore,
the dust detectors aboard
the {\it Ulysses} and {\it Galileo}
spacecrafts
have detected 
large interstellar grains with radii
up to $\simali$2.0$\mum$
flowing through the heliosphere
from the local interstellar cloud
(Gr\"un et al.\ 1994, Kr\"uger et al.\ 2007).
Moreover, even {\it larger} interstellar grains
(of radii $\simali$20$\mum$)
were detected as radar meteors
entering the Earth's atmosphere
on solar-hyperbolic trajectories
(Taylor et al.\ 1996, Baggaley 2000).
More recently, Westphal et al.\ (2014)
reported the detection of seven grains
possibly of interstellar origin
returned by the {\it Stardust} spacecraft
(also see Sterken et al.\ 2015).
These grains are mostly composed
of Mg-rich silicates 
and three of them have radii $\simgt1\mum$.

In addition, Witt et al.\ (2001)
pointed out that the presence of
very large grains in the ISM could be inferred from
X-ray halos
since the X-ray scattering efficiency
varies approximately as $a^{4}$ 
while the differential cross section
varies as $a^{6}$. 
They examined the X-ray halo
around Nova Cygni 1992 and found that,
in order to explain the observed profile and
intensity, 
the grain size distribution needs to extend
to and possibly beyond $\simali$2$\mum$.
Furthermore, surprisingly high-albedo scattering
at near-IR wavelengths has been reported
for several dense regions
(Witt et al. 1994, Block et al.\ 1994,
Lehtinen \& Mattila 1996),
suggesting the possible presence of
a population of grains
at least $\simali$0.5$\mum$ in radii.
Moreover, based on the unaccounted for
O/H abundance (Jenkins 2009, Whittet 2010),
Jenkins (2009), Poteet et al.\ (2015),
and Wang, Li \& Jiang (2015)
argued that the missing reservoir of O
could reside on $\mu$m-sized H$_2$O grains.
It is worth noting that four decades ago
Greenberg (1974) had already pointed out that
there could exist very large grains
comprised of O, C, and N
in the interstellar space.
Socrates \& Draine (2009) suggested that
``pebble''-sized grains of $a$$\simali$1\,mm
may be detected through optical scattered
light halos.
Finally, we note that
$\mu$m-sized grains
were detected in dense cloud cores through
the scattering of the interstellar radiation
at the {\it Spitzer}/IRAC 3.6 and 4.5$\mum$ bands
(i.e., ``coreshine''; Pagani et al.\ 2010,
Steinacker et al.\ 2014).\footnote{%
    The presence of this type of grains
    in dense environments is typically interpreted
    as due to grain growth which also occurs in
    proto-planetary disks
    (e.g., see Bouwman et al.\ 2001,
    Kessler-Silacci et al.\ 2006,
    Ricci et al.\ 2010).
    In contrast, those $\mu$m-sized grains
    discussed in this work are
    for the diffuse ISM as well,
    not confined to dense clouds.
    }

In this work, we aim at constraining
the quantity and size distribution of
the $\mu$m-sized dust population.
This is achieved by fitting the observed
mid-IR extinction at $\simali$3--8$\mum$
(\S\ref{sec:irext})
in terms of the silicate-graphite-PAH model
together with an extra population of $\mu$m-sized grains
(\S\ref{sec:model}).
We present the results in \S\ref{sec:results}
and discuss their astrophysical implications
in \S\ref{sec:discussion}.
Our principal conclusions are
summarized in \S\ref{sec:sum}.

\begin{figure}[!ht]
\centering
\vspace{-0.0in}
\hspace{-2.0in}
\includegraphics[angle=0,width=8.0in]{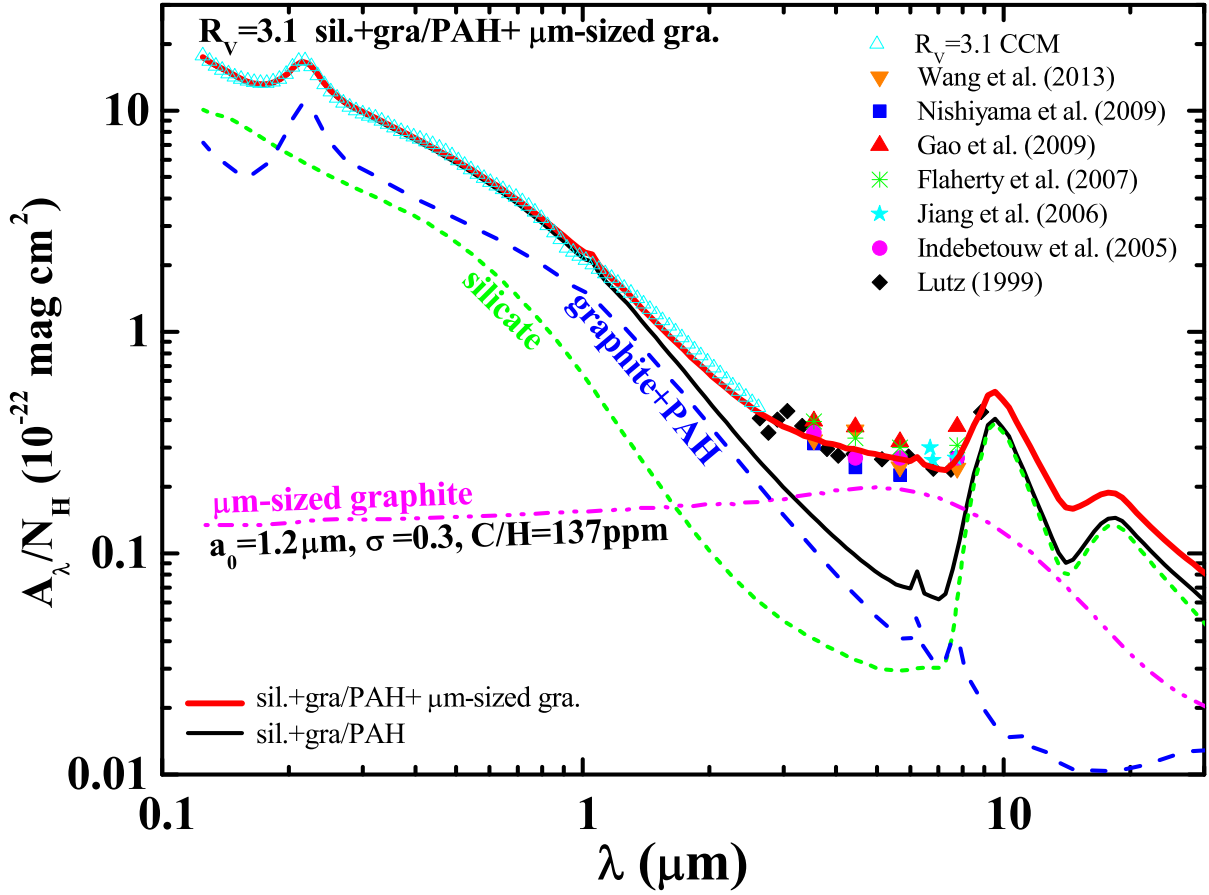}
\hspace{-2.0in}
\vspace{-0.2in}
\caption{\footnotesize
          \label{fig:ext}
          Fitting the $\RV=3.1$ extinction curve from
          the UV/optical to the near- and mid-IR with
          (1) amorphous silicate
          (green short-dashed line),
          (2) graphite and PAHs (blue dashed line), and
          (3) $\mu$m-sized 
          graphite (magenta dash-dot-dotted line)
          with a log-normal size distribution
          characterized by $a_0\approx1.2\mum$,
          $\sigma\approx0.3$, and $\bVLG\approx137\ppm$
          (see eq.\,\ref{eq:VLGdnda}).
          The thick red solid line plots the model-fit
          which is the sum of silicate, graphite/PAHs,
          and $\mu$m-sized graphite.
          The black solid line plots the sum of silicate
          and graphite/PAHs.
          The symbols plot the observed extinction:
          cyan open triangles plot the $\RV=3.1$
          UV/optical/near-IR extinction,
          while other symbols
          plot the mid-IR extinction (see text).
          }
\end{figure}

%
\section{Mid-IR Extinction\label{sec:irext}}
The UV/optical interstellar extinction
can be characterized by
a single parameter $R_V$
(Cardelli et al.\ 1989, CCM).\footnote{%
  $R_V\,\equiv\,A_V/\left(A_B-A_V\right)$
  is the total-to-selective extinction ratio,
  where $A_B$ is
  the blue band extinction
  at $\lambda_B=4400\Angstrom$.
  For the Galactic average, $R_V\approx3.1$.
  }
The UV/optical extinction can be closely fitted
in terms of the classical silicate-graphite model
(Mathis et al.\ 1977, Draine \& Lee 1984 [DL84]).
In the IR at $1\mum < \lambda < 7\mum$,
this model predicts a power-law extinction curve of
$A_\lambda \propto \lambda^{-1.75}$ (Draine 1989).
As elaborated in Wang, Li \& Jiang (2014),
this is too steep to be consistent
with the subsequent observations made by
the {\it Infrared Space Observatory} (ISO)
and the {\it Spitzer Space Telescope}.
Numerous observations suggest that
the mid-IR extinction at $3\mum <\lambda< 8\mum$
is flat or ``gray'' for both diffuse
and dense environments (see Figure~\ref{fig:ext}),
including the Galactic center
(Lutz 1999, Nishiyama et al.\ 2009), 
the Galactic plane
(Indebetouw et al.\ 2005,
Jiang et al.\ 2006, Gao et al.\ 2009),
the Coalsack nebula (Wang et al. 2013),
and nearby star-forming regions (Flaherty et al.\ 2007).
All these observations appear to suggest
an ``universally'' flat mid-IR extinction law,
with little dependence on environments.

\begin{figure}[!ht]
\centering
\vspace{-0.0in}
\hspace{-2.0in}
\includegraphics[angle=0,width=8.0in]{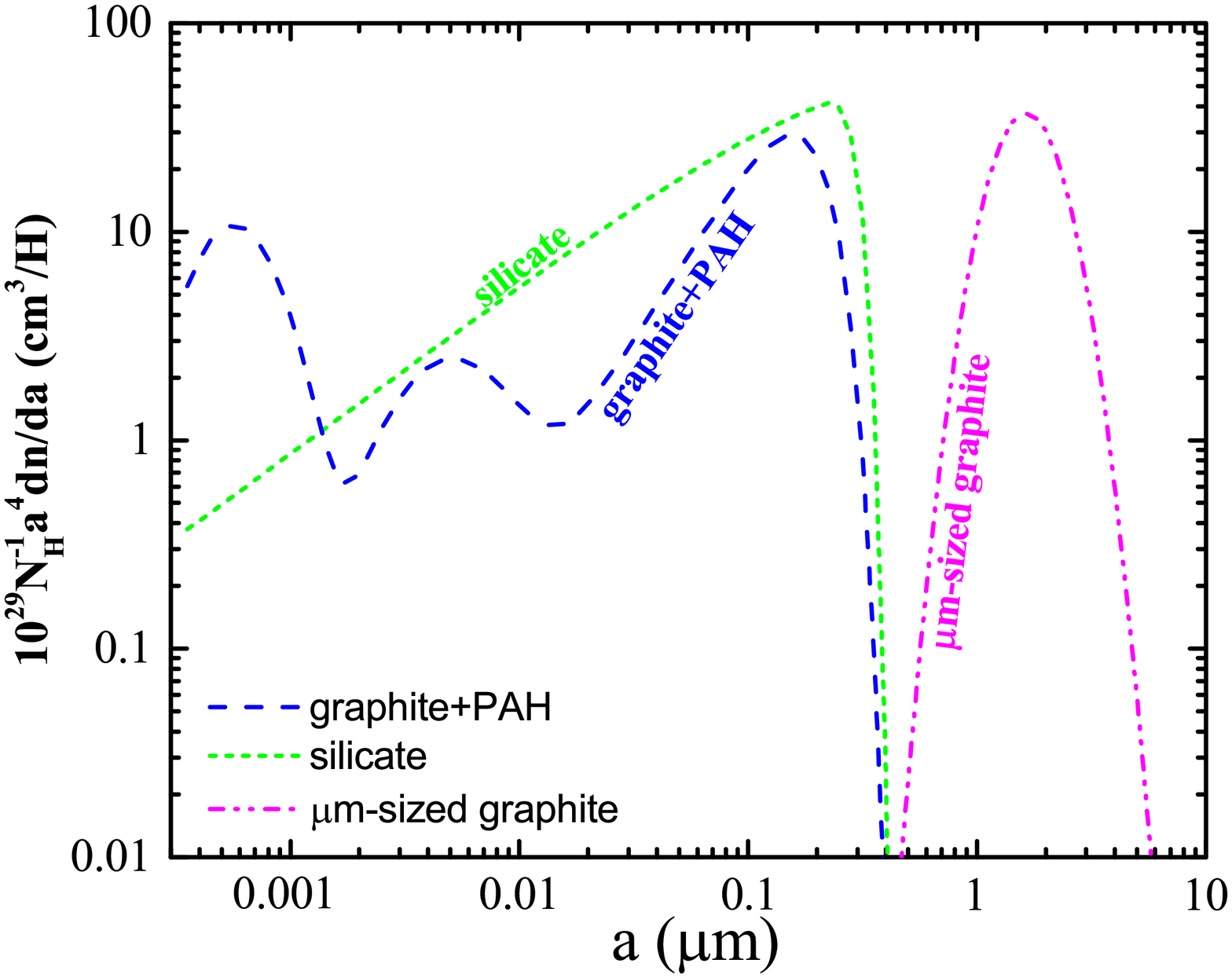}
\hspace{-2.0in}
\vspace{-0.2in}
\caption{\footnotesize
         \label{fig:dnda}
          Grain size distributions for
          (1) silicate
          (green short-dashed line),
          (2) graphite and PAHs (blue dashed line), and
          (3) $\mu$m-sized ``very large'' graphite
          (magenta dash-dot-dotted line).
          }
\end{figure}

%
\section{Dust Model\label{sec:model}}
We aim at reproducing the observed extinction
from the UV/optical to the near- and mid-IR.
We assume a mixture of amorphous silicate dust
and carbonaceous dust, taking the size distribution
functional form of Weingartner \& Draine (2001; WD01).
We assume that the latter extends
from grains with graphitic properties
at radii $a\gtsim50\Angstrom$,
down to grains with PAH-like properties
at very small sizes (LD01).
The WD01 model employs two log-normal size distributions
for two populations of PAHs which respectively
peak at $a_{0,1}$, $a_{0,2}$
and have a width of $\sigma_{1}$, $\sigma_{2}$,
consuming a C abundance of
$b_{{\rm C},1}$, $b_{{\rm C},2}$ (per H nuclei).
Following DL07,
we adopt $a_{0,1}=3.5\Angstrom$, $\sigma_1=0.40$,
$b_{\rm C,1}=45\ppm$, $a_{0,2}=20\Angstrom$,
$\sigma_2=0.55$, and $b_{\rm C,2}=15\ppm$.
These parameters were constrained
by the observed near- and mid-IR emission.
To account for the mid-IR extinction,
we invoke an extra population of very large grains
for which we also adopt
a log-normal size distribution
of peak size $a_0$ and width $\sigma$:
\begin{equation}\label{eq:VLGdnda}
\frac{1}{\nH} \frac{dn}{da}=
\frac{3}{(2\pi)^{3/2}}
\times
\frac{\exp\left(-4.5\sigma^2\right)}{\rho a_0^3 \sigma}
\times
\frac{\bVLG\mu\mH}{2}
\times
\frac{1}{a}\exp\left\{-\frac{1}{2}
\left[\frac{\ln(a/a_0)}{\sigma}\right]^2\right\} ~~,
\end{equation}
where $\mH$ is the atomic H mass,
$\rho$ and $\mu$ are respectively the mass density
and the molecular weight of the dust species
($\rho\approx3.5\g\cm^{-3}$ and $\mu\approx172$ for silicate,
$\rho\approx2.24\g\cm^{-3}$ and $\mu \approx12$ for graphite),
and $\bVLG$ is the abundance per H nuclei locked up
in $\mu$m-sized grains.

We consider 120 wavelengths,
equally spaced in $\ln\lambda$,
to model the extinction between
0.125$\mum$ and 8$\mum$. 
For the ``observed'' extinction
in the wavelength range of
$0.125\mum < \lambda < 3\mum$,
we take the Galactic average of
$R_V=3.1$ as parameterized by CCM.
For the $\simali$3--8$\mum$ mid-IR extinction,
we first obtain a weighted ``average'' 
from the observed extinction shown in Figure~\ref{fig:ext},
with twice as much weight given to
the diffuse sightlines toward
the Galactic center.
We then interpolate the ``average'' mid-IR extinction
into 25 logarithmically equally-spaced wavelengths.

We take the optical constants of
astronomical silicate and graphite of DL84.
The PAH absorption cross sections
are taken from DL07.
Following WD01,
we use the Levenberg-Marquardt method
to minimize $\chi^2 = \chi_1^2 + \chi_V^2$,
where $\chi_1^2$ gives the error
in the extinction fit:
\begin{equation}
\chi_1^2 = \sum_i \frac{\left[\ln A_{\rm obs}(\lambda_i)
       - \ln A_{\rm mod}(\lambda_i) \right]^2}
         {\sigma_i^2}~~~,
\end{equation}
where $A_{\rm obs}(\lambda_i)$
and $A_{\rm mod}(\lambda_i)$ are respectively
the observed and model-computed extinction
at wavelength $\lambda_i$.
Following WD01,
we take the weights $\sigma_i^{-1} = 1$
for $0.125\mum<\lambda<0.9\mum$
and $\sigma_i^{-1} = 1/3$ for
$0.9\mum<\lambda<8\mum$.
The ``penalty'' term $\chi_V^2$ prevents
the model-consumed C and Si abundances
($\cdust$, $\sidust$) from
grossly exceeding their interstellar abundances:
\begin{equation}
\chi_V^2 = 0.4 [\max(\tilde{\rm C},1) -1]^{1.5} +
0.4 [\max(\tilde{\rm Si},1) -1]^{1.5}
\label{eq:chi_V}~~~,
\end{equation}
where $\tilde{\rm C} = \cdust/233\ppm$ and
$\tilde{\rm Si} = \sidust/36.3\ppm$
(ppm is an abbreviation of parts per million).

\section{Results\label{sec:results}}
In fitting the far-UV to mid-IR extinction
represented by 120 ``data points''
(see \S\ref{sec:model}),
we have ten parameters from the WD01 size
distribution function
($C_{\rm g}$, $a_{\rm t,g}$, $a_{\rm c,g}$,
$\alpha_{\rm g}$, $\beta_{\rm g}$ for graphite,
$C_{\rm s}$, $a_{\rm t,s}$, $a_{\rm c,s}$,
$\alpha_{\rm s}$, $\beta_{\rm s}$ for silicate)
and three parameters from the $\mu$m-sized grain
component. Following WD01,
we fix $\acs=0.1\mum$.\footnote{%
  WD01 found that the fitting error only mildly
  varies with $\acs$ provided $\acs\simlt0.1\mum$
  while it abruptly increases with $\acs$
  when $\acs>0.1\mum$.
  }
We consider a graphite composition
for the $\mu$m-sized grain component.\footnote{%
  In the following, unless otherwise stated,
  we refer ``graphite/PAHs'' to nano- and
  sub-$\mu$m-sized graphitic grains
  (including PAHs).
  }
As shown in Figure~\ref{fig:ext},
with $a_0\approx1.2\mum$,
$\sigma\approx0.3$,
and $\bVLG\approx137\ppm$,
the observed extinction from the far-UV
to the mid-IR is closely reproduced
(with $\chi^2\approx0.23$).
The corresponding parameters for the WD01
size distribution function are:
$\Cg\approx5.75\times10^{-12}$,
$\alphag\approx-1.40$,
$\betag\approx0.0291$,
$\atg\approx0.00818\mum$,
and $\acg\approx0.173\mum$ for graphite,
$\Cs\approx7.56\times10^{-14}$,
$\alphas\approx-2.19$,
$\betas\approx-0.586$, and
$\ats\approx0.204\mum$ for silicate.
The derived size distributions are
illustrated in Figure~\ref{fig:dnda}.

The model requires a total silicate mass
(relative to H)
of $M_{\rm sil}/M_{\rm H}\approx6.95\times10^{-3}$,
and a total carbonaceous dust mass
(relative to H)
of $M_{\rm carb}/M_{\rm H}\approx4.34\times10^{-3}$,
indicating a total gas-to-dust mass ratio
of $M_{\rm gas}/M_{\rm dust}\approx124$,
where $M_{\rm gas}\approx M_{\rm He} + M_{\rm H}
\approx 1.4\,M_{\rm H}$ for ${\rm He/H\approx0.1}$.
The required elemental depletions are
$\sidust\approx40.4\ppm$ and
$\cdust\approx362\ppm$.
The latter includes
$\cVLG\approx137\ppm$ in $\mu$m-sized graphite grains
and $\cPAH=60\ppm$ in PAHs,
which respectively account
for $\simali$37.8\% and $\simali$16.6\%
of the total mass of
the carbonaceous dust component,
and $\simali$14.6\% and $\simali$6.38\%
of the total dust mass.

We have also tried $\mu$m-sized silicate grains.
Compared with graphite, silicates are
much more transparent at $\lambda<8\mum$.
To account for the observed mid-IR extinction,
one requires $\siVLG\approx2.95\times10^4\ppm$
to be locked up in $\mu$m-sized silicates,
far exceeding the available amount
of $\siism$\,$\approx$\,32--41$\ppm$ in the ISM
(see \S\ref{sec:discussion}).
Also, $\mu$m-sized silicate grains exhibit
a Si--O resonance dip at $\simali$8$\mum$
which is not seen in the observed extinction curve.
We have also considered $\mu$m-sized iron grains.
Unless they are extremely-elongated
like needles (Dwek 2004), the mid-IR extinction
requires a depletion of
$\feVLG\approx100\ppm$,
far exceeding the available amount
of Fe/H\,$\approx$\,28--35$\ppm$ in the ISM.
More recently, K\"ohler et al.\ (2014) found that
amorphous silicates with iron nanoparticles embedded
could enhance the mid-IR extinction.
K\"ohler et al.\ (2015) further found that dust aggregates
could also cause increased mid-IR extinction.
It would be interesting to see how their model extinction
per H column compares with the observed extinction.

\begin{figure}[!ht]
\centering
\vspace{-0.0in}
\hspace{-2.0in}
\includegraphics[angle=0,width=4.0in]{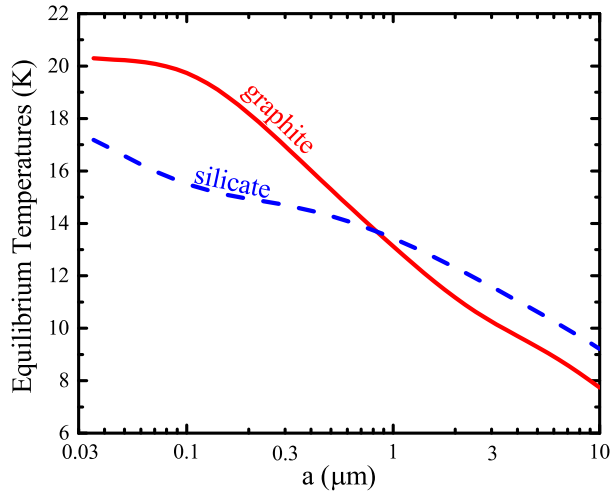}
\hspace{-2.0in}
\vspace{-0.2in}
\caption{\footnotesize
        \label{fig:Teq}
	Equilibrium temperatures for silicate (dashed line)
        and graphite grains (solid line) heated by
        the MMP83 ISRF.
        }
\end{figure}

\begin{figure}[!ht]
\centering
\vspace{-0.0in}
\hspace{-2.0in}
\includegraphics[angle=0,width=8.0in]{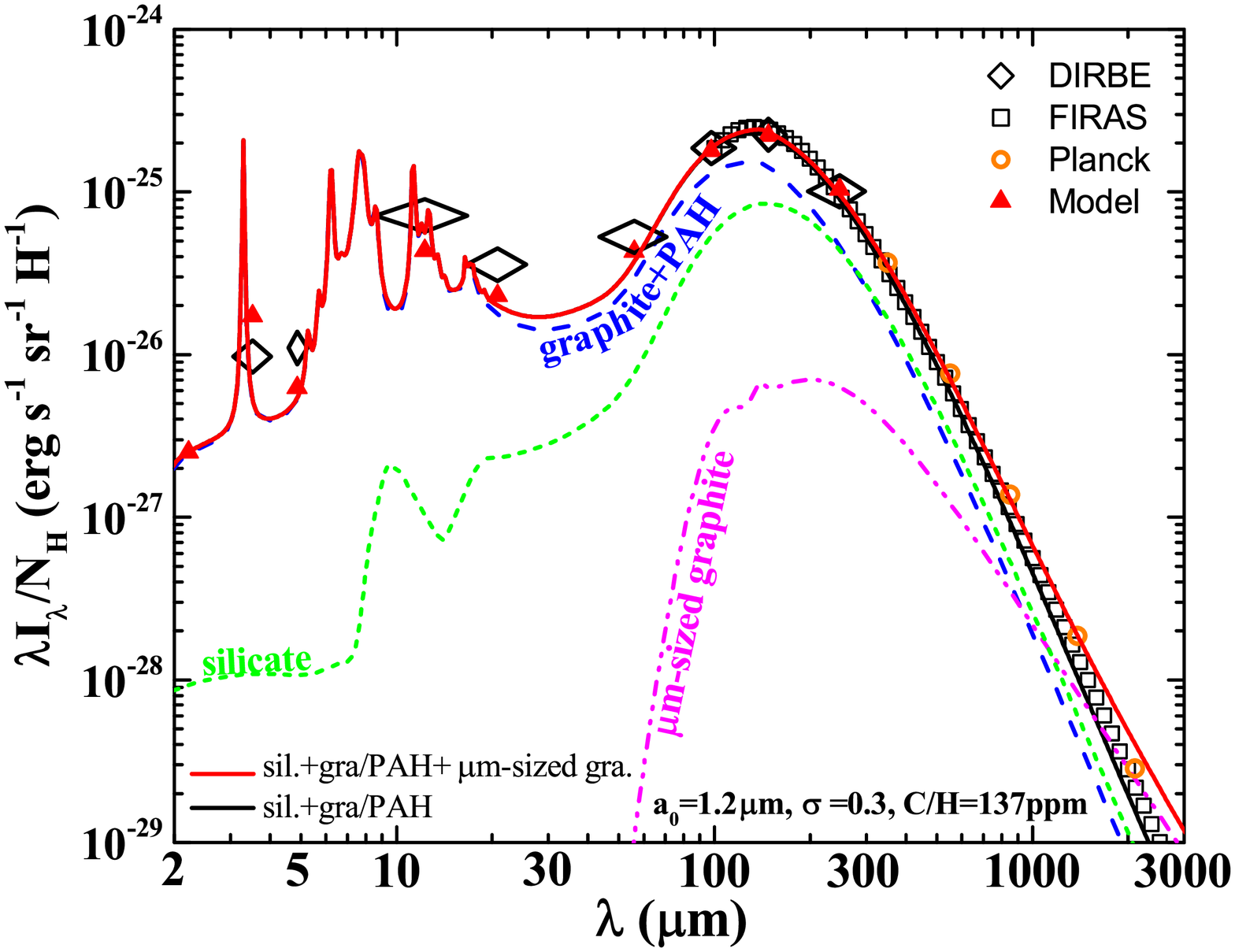}
\hspace{-2.0in}
\vspace{-0.2in}
\caption{\footnotesize
         \label{fig:irem}
          Comparison of the model to
          the observed emission
          from the diffuse ISM.
          Green short-dashed line shows emission
          from amorphous silicate;
          blue dashed line shows emission
          from graphite/PAHs;
          magenta dash-dot-dotted line shows
          emission from $\mu$m-sized graphite grains.
          Red triangles show the model spectrum
          (red solid line)
          convolved with the DIRBE filters.
          Observational data are from DIRBE
          (black diamonds; Arendt et al.\ 1998),
          FIRAS (black squares; Finkbeiner et al.\ 1999),
          and Planck (orange circles;
          Planck Collaboration XVII 2014).
          }
\end{figure}

%
\section{Discussion\label{sec:discussion}}
\subsection{IR Emission\label{sec:irem}}
As shown in Figure~\ref{fig:ext},
$\mu$m-sized graphite grains are
``gray'' in the UV, optical, and near-IR.
They do not absorb much in the UV/optical,
and therefore by implication do not emit
much radiation in the IR.
We have calculated the temperature probability
distribution functions
of PAHs, small graphite and silicate grains
of radii smaller than $\simali$250$\Angstrom$
which are heated by
the Mathis, Mezger, \& Panagia (1983, MMP83)
interstellar radiation field (ISRF).
We have also calculated the equilibrium temperatures
of silicate and graphite grains of radii
larger than $\simali$250$\Angstrom$
(see Figure~\ref{fig:Teq}).
The resulting IR emission is shown in
Figure~\ref{fig:irem} and compared with
that of the diffuse ISM observed
by {\it COBE}/DIRBE,
{\it COBE}/FIRAS and {\it Planck}.\footnote{%
   Many {\it Herschel} observations have deen
   dedicated to the investigation of
   the dust properties in the ISM
   (e.g., see Abergel et al.\ 2010, 
   Gordon et al.\ 2010, Molinari et al.\ 2010, 
   Juvela et al.\ 2011, Juvela 2015).
   In this work, these {\it Herschel} data
   are not included
   since the {\it Herschel}/PACS bands
   are essentially covered by {\it COBE}/DIRBE,
   while the {\it Herschel}/SPIRE bands
   are essentially covered by {\it COBE}/FIRAS
   and {\it Planck}
   (see Table~4 of LD01 and Table~6 of DL07).
   }
Our model yields a total IR intensity of
$\simali$$4.55\times10^{-24}\erg\s^{-1}\rmH^{-1}$.
The fractional contributions of
silicate, graphite/PAHs, and $\mu$m-sized graphite
are approximately 27.5\%, 70.0\%, and 2.5\%, respectively.
The $\mu$m-sized graphite component is cold
($T\approx13.1, 12.5, 11.2\K$
for $a=1, 1.2, 2\mum$ compared to
$T\approx20\K$ for $a=0.1\mum$, see Figure~\ref{fig:Teq})
and its contribution to the overall emission
is only noticeable at $\lambda\simgt550\mum$,
becomes significant at $\lambda\gtsim1382\mum$,
and dominates the emission at $\lambda\gtsim2098\mum$.

Figure~\ref{fig:irem} shows that our model fits
the observed emission very well
from the near-IR to submm:
up to $\lambda\simlt850\mum$,
it is in close agreement
with the {\it COBE}/FIRAS spectrophotometry.
At $\lambda>850\mum$, the model emission
slightly exceeds the {\it COBE}/FIRAS data.
We note that the {\it Planck} photometric data
at $\lambda\gtsim850\mum$ are also
slightly higher than the {\it COBE}/FIRAS
spectrophotometry.
Our model is in excellent agreement with
the {\it Planck} data up to $\lambda\simgt1382\mum$,
and does not overpredict until $\lambda\simgt2098\mum$.
At $\lambda=3000\mum$, the model emission
is stronger than the {\it COBE}/FIRAS photometry
by a factor of $\simali$1.5 and exceeds the
{\it Planck} data by $\simali$72\%.
However, the far-IR dielectric functions of
graphite and silicate adopted here may not be
known with a high precision
(e.g., at $\lambda=1000\mum$ the DL84 silicate
has an opacity of $\kappa_{\rm abs}\approx0.33\cm^2\g^{-1}$,
while the silicate material measured by
Agladze et al.\ (1996) had
$\kappa_{\rm abs}\approx1.25\cm^2\g^{-1}$).
We also note that the {\it Planck} data
at $\lambda=3000\mum$ exceeds that of
{\it COBE}/FIRAS by $\simali$45\%.

Finkbeiner, Davis \& Schlegel (1999; FDS)
approximated the {\it COBE}/FIRAS data
in terms of a two-component model
consisting of a warm component of
temperature $T_W\approx16.2\K$
and a cold component of
temperature $T_C\approx9.4\K$:
$I_\lambda \sim \lambda^{-2.70} B_\lambda(T_W)
+ 0.47\,\lambda^{-1.67} B_\lambda(T_C)$.
They ascribed the warm component to
carbonaceous grains and the cold component
to amorphous silicates.
We argue that while the FDS two-component model
provides an excellent representation of
the observed far-IR emission,
their ascription may not be physical.
As shown in Figure~\ref{fig:Teq},
to attain an equilibrium temperature
of $T_C=9.4\K$, silicate grains need to
be very large, with radii $a\approx8.9\mum$.
Such very large silicate grains are gray
in the UV/optical and contribute little
to the UV/optical extinction.
As a result, the observed UV/optical extinction
would exclusively rely on carbonaceous dust.
However, it is known that neither silicates
nor carbonaceous dust \emph{alone} can account for
the observed UV/optical extinction (see Li 2005a).
It is more natural to attribute the warm component
to a combination of sub-$\mu$m-sized silicate and
carbonaceous grains and the cold component
either to a population of $\mu$m-sized grains
or alternatively, according to
the two-level-system (TLS) model,
to the low energy transitions associated with
the disordered internal structure
of sub-$\mu$m-sized amorphous grains
(Meny et al.\ 2007).

A cold emission component of $T$\,$\simali$4--7$\K$
was also noted in the {\it COBE}/FIRAS
Galactic emission spectrum
by Wright et al.\ (1991)
and Reach et al.\ (1995).
Rowan-Robinson (1992) explained the cold emission
in terms of grains of $a=30\mum$.
But he did not specify the composition of
this grain population. He simply assumed
its absorption efficiency to be
$Q_{\rm abs}(a,\lambda) = 1$
for $\lambda < 4\pi a$
and $Q_{\rm abs}(a,\lambda) = 4\pi a/\lambda$
for $\lambda \ge 4\pi a$.

As shown in Figure~\ref{fig:irem},
our model SED flattens
at $\lambda\gtsim1382\mum$.
The overall dust opacity at $\lambda<1382\mum$
has a power-index of $\beta_{\rm FIR}\approx1.89$.
In contrast, the opacity at $\lambda>1382\mum$
has a slightly smaller power-index of
$\beta_{\rm mm}\approx1.67$.
The {\it Planck} observations of
the Galactic diffuse ISM detected
the flattening of the SED at long wavelengths,
with $\beta_{\rm FIR} - \beta_{\rm mm} \approx0.15$
at $\lambda > 850\mum$
(see Planck Collaboration XVII 2014).
These observations also provided evidence
for the far-IR opacity variations
in the diffuse ISM, attributable to
large amorphous carbon grains
or the TLS model of disordered
charge distribution in amorphous grains.

\subsection{The C/H Crisis\label{sec:C/H}}
The interstellar abundances of Si/H and C/H are unknown.
They are often assumed to be that of
solar ($\sisun\approx32.4\pm2.2\ppm$,
$\csun\approx269\pm31\ppm$, Asplund et al.\ 2009),
proto-Sun ($\sisun\approx40.7\pm1.9\ppm$,
$\csun\approx288\pm27\ppm$, Lodders 2003),
early B stars ($\sistar\approx31.6\pm3.6\ppm$,
$\cstar\approx214\pm20\ppm$, Nieva \& Przybilla 2012),
or young F/G stars
($\sistar\approx39.9\pm13.1\ppm$,
$\cstar\approx358\pm82\ppm$, Sofia \& Meyer 2001).
Our model consumes $\sidust\approx40.4\ppm$
which is consistent with the proto-Sun Si/H abundance.
With the gas-phase abundance
of $\cgas\approx140\ppm$ (Cardelli et al.\ 1996)
or $\cgas\approx100\ppm$ (Sofia et al.\ 2011) subtracted,
the proto-Sun reference standard leaves
only $\simali$148$\ppm$ or $\simali$188$\ppm$
of C/H for the dust, respectively.
With $\cdust\approx362\ppm$,
our model uses too much C
compared to what would be
available in the ISM.
Even ignoring the $\mu$m-sized graphite component,
the silicate-graphite-PAH model,
with $\cdust\approx252\ppm$ (WD01),
is already in ``C crisis''
(Snow \& Witt 1995).
With the inclusion of a population of
$\mu$m-sized graphite grains,
our model amplifies
the so-called ``C crisis''.

The ``C crisis'' holds for
all dust models, including that of
Zubko et al.\ (2004; ZDA)
which requires $\cdust\approx244\ppm$
(and $\sidust\approx36\ppm$)
and Jones et al.\ (2013; J13)
which requires $\cdust\approx233\ppm$
(and $\sidust\approx50\ppm$).
Note that the mid-IR extinction
has not been accounted for by ZDA or J13.
The ZDA model produces much less extinction
at $\lambda > 1\mum$ than observed
(see Figure~23.11 in Draine 2011).
The J13 model fitted the IR extinction
tabulated in Mathis (1990) which is much
lower than the mid-IR extinction described
here in \S\ref{sec:irext}.
The total consumed C/H is expected to
be similar to ours if the flat mid-IR
extinction is to be accounted for by
their models.
This is indicated by
the Kramers-Kronig relation
of Purcell (1969)
which relates the wavelength-integrated extinction
to the dust quantity:
$\int_{0}^{\infty} A_\lambda/\NH\,d\lambda
= 1.086\times 3 \pi^2 F \left(V_{\rm dust}/{\rm H}\right)$
where $V_{\rm dust}/{\rm H}$ is the total dust volume
per H nucleon,
and $F$ is a dimensionless factor
which depends only upon the grain shape and
the static dielectric constant
of the grain material.
Spherical silicates ($F\approx0.64$) with
the proto-Sun abundance
contribute $\approx$$6.53\times10^{-26}
\magni\cm^{3}\rmH^{-1}$ to the extinction integration.
With $\int_{912\Angstrom}^{1000\mum}
A_\lambda/\NH\,d\lambda \approx1.70\times10^{-25}
\magni\cm^{3}\rmH^{-1}$
(see Figure~\ref{fig:ext}),
one derives $\cdust\approx365\ppm$
for spherical graphite ($F\approx1.0$).
This is a lower limit since
$\int_{0}^{\infty} A_\lambda/\NH\,d\lambda
> \int_{912\Angstrom}^{1000\mum}
A_\lambda/\NH\,d\lambda$.
For moderatly elongated grains,
the C crisis still persists
(e.g., with an elongation of $\simali$2--3,
one requires $\cdust\approx275\ppm$
of graphite of $F\approx1.25$).

A likely solution to the C crisis problem
is that the stellar photospheric abundances
of dust-forming elements may be considerably
lower than that of the interstellar
material from which young stars are formed.
This could be caused by an incomplete incorporation
of heavy elements in stars
during the star formation process,
and/or an underestimation of
the degree of heavy-element settling
in stellar atmospheres (see Li 2005b).
Parvathi et al.\ (2012) derived
the gas-phase C/H abundance of
16 Galactic interstellar sightlines.
They found that $\cgas$ varies from one sightline
to another, with $\simali$1/3 of the sightlines
having their gas-phase C/H abundance {\it alone}
exceeding the proto-Sun C/H abundance 
(e.g., the sightline toward HD\,206773
has a gas-phase C/H abundance of
$\approx464\pm57\ppm$ [see Parvathi et al.\ 2012]
which is comparable to
the total C/H abundance
of $\simali$462$\ppm$
required by our model
(i.e., $\cdust\approx362\ppm$
[see \S\ref{sec:results}]
plus $\cgas\approx100\ppm$
[see Sofia et al.\ 2011]).\footnote{%
  If we attribute the flat $\simali$3--8$\mum$
  mid-IR extinction to $\mu$m-sized silicate or
  iron spheres,  we would face a much more
  severe Si or Fe ``crisis'':
  as discussed in \S\ref{sec:results},
  while the model consisting of
  $\mu$m-sized graphite grains
  requires $\simali$60\% more C/H
  than the proto-Sun C/H abundance,
  the model consisting of
  $\mu$m-sized silicate grains
  requires $\simali$725 times more Si/H
  than the proto-Sun Si/H abundance,
  and the model consisting of
  $\mu$m-sized iron spheres
  requires $\simali$3.0 times more Fe/H
  than the proto-Sun Fe/H abundance.
  }

\subsection{An Artificial Gap
            in the Grain Size Distribution?
            \label{sec:sizegap}}
The size distribution derived for
the $\mu$m-sized dust component
is not a continuous extension
of the sub-$\mu$m-sized component
but has a gap in between
(see Figure~\ref{fig:dnda}).
This seems inconsistent
with the observed continuous size distribution
of the interstellar grains
entering the solar system
(see Figure~2 of Frisch et al.\ 1999).
However, the extinction resulting from
such a continuous size distribution
would be totally unlike
what is actually observed
(see Figures~3,\,4 of Draine 2009).\footnote{%
   Draine (2009) added a population of large grains
   to the WD01 size distribution so that the new one
   approximately reproduces that of the interstellar
   grains detected by {\it Ulysses} and {\it Galileo}
   at $m > 3\times10^{-13}\g$ of Landgraf et al.\ (2000).
   With the inclusion of these large grains,
   Draine (2009) derived an extinction curve
   of $R_V\approx5.8$.
   }
As a matter of fact, a close inspection
of the {\it in situ} dust size distribution
of Frisch et al.\ (1999) 
reveals that it already shows 
a flattening off at $a\gtsim0.5\mum$.
Witt et al.\ (2001) also noted that
the size distribution continuously
extending to micrometers required
to explain the X-ray halo around
Nova Cygni 1992
would lead to an extinction curve
with $R_V \approx 6.1$.
It would be interesting to see
if a size distribution with a gap
in between the sub-$\mu$m-sized grains
and the $\mu$m-sized grains
like the one derived here
could reproduce the observed X-ray halo
and result in an extinction curve
with $R_V \approx 3.1$.

\subsection{The Origin of Very Large Grains\label{sec:origin}}
The origin of
$\mu$m-sized interstellar grains is not known.
Theoretical calculations show that
carbon dust can grow to sizes as large as
a few micrometers in the radioactive
environment of supernova ejecta,
even if O is more abundant than C
(Clayton et al. 1999).
These calculations are also applicable to
the formation of $\mu$m-sized oxides
within supernova gas having C\,$>$\,O.
However, we do not know in the ISM
how much dust originates from supernovae.
If a substantial fraction of interstellar dust
is from supernova condensates,
then $\mu$m-sized grains
may be prevalent in the ISM.

We note that the size
and quantity derived for
the $\mu$m-sized graphite dust
does not rely on
the silicate-graphite-PAH model.
The $\mu$m-sized component
is inferred from the mid-IR extinction
which is essentially independent
of the observed properties
(e.g., UV/optical/near-IR extinction,
near-, mid-, and far-IR emission)
previously used to constrain dust models.
These models differ from each other
in detailed composition and morphology
(see Li 2004). But they all fit
the UV/optical extinction and are
deficient in comparison with
the observed $\simali$3--8$\mum$
mid-IR extinction
except the WD01 $R_V=5.5$ model.
To account for the observed mid-IR
extinction, a population of $\mu$m-sized
dust is required by all dust models.
The size is inferred from
a general consideration of
light scattering theory.
The quantity of this component
can be inferred from
the Kramers-Kronig relation
of Purcell (1969).

Finally, we note that,
while the $\mu$m-sized dust component
derived here is carbonaceous in nature,
the $\mu$m-sized interstellar grains
detected by the {\it Stardust} spacecraft
are remarkably devoid of carbon
(Westphal et al.\ 2014).
This is probably due to the
shock destruction of carbonaceous grains
in the local interstellar cloud,
as revealed by the overabundant CII
in the interstellar gas
surrounding the heliosphere
(see Slavin \& Frisch 2008).
\section{Summary\label{sec:sum}}
The $\simali$3--8$\mum$ mid-IR extinction
is observed to be flat or gray in various
interstellar environments,
including low-density diffuse clouds,
translucent clouds, and dense clouds.
It provides a sensitive constraint
on very large, $\mu$m-sized grains
which are gray at optical wavelengths
and whose existence can not be constrained
by the far-UV to near-IR extinction.
We model the flat $\simali$3--8$\mum$
mid-IR extinction with $\mu$m-sized grains
together with a mixture of
silicate and graphite grains of sizes
ranging from a few angstroms
to a few submicrometers.
Our principal results are as follows:
\begin{enumerate}
\item The observed interstellar extinction is
      closely reproduced from the far-UV to
      the mid-IR with $\sidust\approx40.4\ppm$ in
      silicate grains,
      $\cdust\approx362\ppm$
      in carbonaceous grains
      which include PAHs ($\cPAH\approx60\ppm$),
      submicron-sized graphite
      (C/H\,$\approx$\,165$\ppm$), and
      $\mu$m-sized graphite ($\cVLG\approx137\ppm$).
      The sizes of the $\mu$m-sized graphite component
      are modeled as a log-normal distribution
      peaking at $a_0\approx1.2\mum$
      and having a width of $\sigma\approx0.3$.
\item Our model closely reproduces the IR emission
      of the diffuse ISM observed by {\it COBE}/DIRBE,
      {\it COBE}/FIRAS, and {\it Planck}
      from the near-IR up to $\lambda\simgt1382\mum$.
      With equilibrium temperatures of $T\simlt13\K$,
      $\mu$m-sized graphite grains
      dominate the emission at $\lambda\simgt2098\mum$
      and account for $\simali$2.5\%
      of the total IR emission.
\item A carbon budget problem does arise
      with respect to our model when one
      compares the model-required C depletion
      of $\cdust\approx362\ppm$
      with the solar and proto-Sun abundances
      of $\csun$\,$\approx$\,269, 288$\ppm$.
\item Micrometer-sized silicate or iron grains,
      if present in the ISM,
      are unlikely responsible for
      the observed $\simali$3--8$\mum$ extinction
      as they would require too much Si/H and Fe/H,
      exceeding the proto-Sun Si/H and Fe/H abundances
      by a factor of $\simali$725 and 3.0, respectively.
\end{enumerate}

\acknowledgments{
We thank A.C.A.~Boogert, J.~Gao,
B.W.~Holwerda, J.Y.~Seok, V.J.~Sterken,
A.N.~Witt, Y.X.~Xie,
and the anonymous referee
for helpful comments/suggestions.
This work is
supported by NSFC 11173007, 11373015,
973 Program 2014CB845702,
NSF AST-1109039, and NNX13AE63G.
S.W. acknowledges support from
the China Scholarship Council
(No.\,201406040138).
}



\begin{thebibliography}{}

\bibitem[]{}Abergel, A., Arab, H., Compi{\`e}gne, M., 
            et al.\ 2010, \aap, 518, L96
\bibitem[]{}Agladze, N. I., Sievers, A. J., Jones, S. A.,
            et al.\ 1996, \apj, 462, 1026
\bibitem[]{}Arendt, R. G., Odegard, N., Weiland, J. L., 
            et al.\ 1998, \apj, 508, 74
\bibitem[]{}Asplund, M., Grevesse, N., Sauval, A.J.,
            \& Scott, P.\  2009, ARA\&A, 47, 481
\bibitem[]{}Baggaley, W. J.\ 2000, \jgr, 105, 10353
\bibitem[]{}Block, D. L., Witt, A. N., Grosbol, P.,
            et al.\ 1994, \aap, 288, 383
\bibitem[]{}Bouwman, J., Meeus, G., de Koter, A., et al.\ 
            2001, \aap, 375, 950
\bibitem[]{}Cardelli, J. A.,Clayton, G. C., \& Mathis, J. S.
            1989, \apj, 345, 245 (CCM)
\bibitem[]{}Cardelli, J. A., Meyer, D. M., Jura, M., 
            \& Savage, B. D.\ 1996, \apj, 467, 334
\bibitem[]{}Clayton, D. D., Liu, W., \& Dalgarno, A.
            1999, Science, 283, 1290
\bibitem[]{}Clayton, D. D., \& Nittler, L. R. 2004,
            \araa, 42, 39
\bibitem[]{}Draine, B. T. 1989,
            in Infrared Spectroscopy in Astronomy,
            ed. B. H. Kaldeich
            (Paris: ESA Publ. Division), 93
\bibitem[]{}Draine, B.~T.\ 2009, Space Sci. Rev.,
            143, 333
\bibitem[]{}Draine, B. T. 2011,
            Physics of the Interstellar and Intergalactic Medium
            (Princeton, NJ: Princeton Univ. Press)
\bibitem[]{}Draine, B. T., \& Lee, H. M. 1984,
            \apj, 285, 89 (DL84)
\bibitem[]{}Draine, B. T., \& Li, A. 2001, \apj,
            551, 807
\bibitem[]{}Draine, B. T., \& Li, A. 2007, \apj,
            657, 810 (DL07)
\bibitem[]{}Dwek, E.\ 2004, \apjl, 611, L109
\bibitem[]{}Flaherty, K. M., Pipher, J. L.,
            Megeath, S. T., et al.\ 2007, \apj, 663, 1069
\bibitem[]{}Finkbeiner, D. P., Davis, M., \&
            Schlegel, D. J. 1999, \apj, 524, 867 (FDS)
\bibitem[]{}Frisch, P. C., Dorschner, J. M.,
            Geiss, J., et al.\ 1999, \apj, 525, 492
\bibitem[]{}Gao, J., Jiang, B.W., \& Li, A., 2009,
            \apj, 707, 89
\bibitem[]{}Gordon, K.~D., Galliano, F., Hony, S., 
            et al.\ 2010, \aap, 518, L89
\bibitem[]{}Greenberg, J. M.\ 1974, \apjl, 189, L81
\bibitem[]{}Gr\"un, E., Gustafson, B., Mann, I., et al.\
            1994, \aap, 286, 915
\bibitem[]{}Indebetouw, R., Mathis, J. S., Babler, B. L., 
            et al.\ 2005, \apj, 619, 931
\bibitem[]{}Jenkins, E. B.\ 2009, \apj, 700, 1299
\bibitem[]{}Jiang, B.W., Gao, J., Omont, A.,
            Schuller, F., \& Simon, G.\ 2006,
            \aap, 446, 551
\bibitem[]{}Jones, A. P., Fanciullo, L., K{\"o}hler, M.,
                  et al.\ 2013, \aap, 558, A62 (J13)
\bibitem[]{}Juvela, M., Ristorcelli, I., Pelkonen, V.-M., 
            et al.\ 2011, \aap, 527, A111
\bibitem[]{}Juvela, M. 2015, \planss, in press            
\bibitem[]{}Kessler-Silacci, J., Augereau, J.-C., 
            Dullemond, C.~P., et al.\ 2006, \apj, 639, 275
\bibitem[]{}K{\"o}hler, M., Jones, A., \& Ysard, N.
            2014, \aap, 565, L9
\bibitem[]{}K{\"o}hler, M., Ysard, N., \& Jones, A.~P.\ 
             2015, \aap, 579, A15 
\bibitem[]{}Kr\"uger, H., Landgraf, M. Altobelli, N.,
            et al.\ 2007, \ssr, 130, 401
\bibitem[]{}Landgraf, M., Baggaley, W. J., Gr{\"u}n, E.,
            Kr{\"u}ger, H., \& Linkert, G.\
            2000, \jgr, 105, 10343
\bibitem[]{}Lehtinen, K., \& Mattila, K.\ 1996,
            \aap, 309, 570
\bibitem[]{}Li, A.\ 2004, in ASP Conf. Ser. 309,
            Astrophysics of Dust,
            ed. A. N. Witt, G. C. Clayton, \& B. T. Draine
            (San Francisco, CA: ASP), 417
\bibitem[]{}Li, A.\ 2005a, J. Phys.: Conf. Ser., 6, 229
\bibitem[]{}Li, A.\ 2005b, \apj, 622, 965
\bibitem[]{}Li, A.\ 2009,
            in Small Bodies in Planetary Sciences,
            ed. I. Mann, A. Nakamura, \& T. Mukai
            (Berlin: Springer), 167
\bibitem[]{}Li, A., \& Draine, B. T. 2001, \apj,
            554, 778 (LD01)
\bibitem[]{}Lodders, K. 2003, \apj, 591, 1220
\bibitem[]{}Lutz, D. 1999,
            in The Universe as Seen by ISO,
            ed. P. Cox \& M. Kessler
            (ESA Special Publ., Vol.~427;
             Noordwijk: ESA), 623
\bibitem[]{}Mathis, J. S., Mezger, P. G., \& Panagia, N.
            1983, \aap, 128, 212 (MMP83)
\bibitem[]{}Mathis, J. S., Rumpl, W., \& Nordsieck, K. H.
            1977, \apj, 217, 425
\bibitem[]{}Meny, C., Gromov, V., Boudet, N., et al.\
            2007, \aap, 468, 171
\bibitem[]{}Molinari, S., Swinyard, B., Bally, J., 
            et al.\ 2010, \pasp, 122, 314
\bibitem[]{}Nieva, M. F., \& Przybilla. N., 2012,
            \aap, 539, 143
\bibitem[]{}Nishiyama, S., Tamura, M., Hatano, H.,
            et al.\ 2009, \apj, 696, 1407
\bibitem[]{}Pagani, L., Steinacker, J., Bacmann, A.,
            et al.\ 2010, Science, 329, 1622
\bibitem[]{}Parvathi, V. S., Sofia, U. J., Murthy, J.,
            \& Babu, B. R. S.\ 2012, \apj, 760, 36
\bibitem[]{}Planck Collaboration XVII 2014,
            \aap, 566, A55
\bibitem[]{}Poteet, C. A., Whittet, D. C. B.,
            \& Draine, B. T.\ 2015, \apj, 801, 110
\bibitem[]{}Purcell, E. M. 1969, \apj, 158, 433
\bibitem[]{}Reach, W. T., Dwek, E., Fixsen, D. J., 
            et al.\ 1995, \apj, 451, 188
\bibitem[]{}Ricci, L., Testi, L., Natta, A., 
            \& Brooks, K.~J.\ 2010, \aap, 521, A66
\bibitem[]{}Rowan-Robinson, M.\ 1992, \mnras, 258, 787
\bibitem[]{}Slavin, J. D., \& Frisch, P. C.\ 2008,
            \aap, 491, 53
\bibitem[]{}Snow, T. P., \& Witt, A. N. 1995,
            Science, 270, 1455
\bibitem[]{}Socrates, A., \& Draine, B. T. 2009, \apjl,
            702, L77
\bibitem[]{}Sofia, U. J., \& Meyer, D.M., 2001,
            \apjl, 554, L221
\bibitem[]{}Sofia, U. J., Parvathi, V. S., Babu, B. R. S., 
            \& Murthy, J.\ 2011, \aj, 141, 22
\bibitem[]{}Steinacker, J., Ormel, C. W., Andersen, M.,
            \& Bacmann, A. 2014, \aap, 564, 96
\bibitem[]{}Sterken, V. J., Strub, P., Kr\"uger, H., 
            et al.\ 2015, \apj, in press
\bibitem[]{}Taylor, A.D., Baggaley, W.J.,
            \& Steel, D.I., 1996, \nat, 380, 323
\bibitem[]{}Trumpler, R. J. 1930, \pasp, 42, 214
\bibitem[]{}Wang, S., Gao, J., Jiang, B. W., Li, A.,
            \& Chen, Y. 2013, \apj, 773, 30
\bibitem[]{}Wang, S., Li, A., \& Jiang, B. W. 2014,
            \planss, 100, 32
\bibitem[]{}Wang, S., Li, A., \& Jiang, B. W. 2015,
            \mnras, in press
\bibitem[]{}Weingartner, J. C., \& Draine, B. T.
            2001, \apj, 548, 296 (WD01)
\bibitem[]{}Westphal, A. J., Stroud, R. M.,
            Bechtel, H. A., et al.\
            2014, Science, 345, 786
\bibitem[]{}Whittet, D. C. B.\ 2010, \apj, 710, 1009
\bibitem[]{}Witt, A. N., Lindell, R. S., Block, D. L.,
            \& Evans, Rh. 1994, \apj, 427, 227
\bibitem[]{}Witt, A. N., Smith, R. K., \& Dwek, E.
            2001, \apjl, 510, L201
\bibitem[]{}Wright, E. L., Mather, J. C., Bennett, C. L.,
            et al.\ 1991, \apj, 381, 200
\bibitem[]{}Zubko, V., Dwek, E., \& Arendt, R. G.
            2004, \apjs, 152, 211 (ZDA)

\end{thebibliography}
\end{document}